\documentclass[12pt,preprint]{aastex}
\usepackage{natbib}
\usepackage{amsmath}

\shorttitle{A Radio-Loud NLS1}
\shortauthors{Oshlack, Webster, Whiting}

\begin{document}

\title{A Very Radio-Loud Narrow-Line Seyfert 1: PKS 2004-447}

\author{A. Y. K. N. Oshlack,}
\affil{School of Physics, University of Melbourne, Parkville, Victoria, Australia, 3010}
\email{aoshlack@physics.unimelb.edu.au}

\author{R. L. Webster}
\affil{School of Physics, University of Melbourne, Parkville, Victoria, Australia, 3010}
\email{rwebster@physics.unimelb.edu.au}

\author{and M. T. Whiting}
\affil{School of Physics, University of Melbourne, Parkville, Victoria, Australia, 3010}
\email{mwhiting@physics.unimelb.edu.au}

\begin{abstract}

We have discovered a very radio-loud Narrow-Line Seyfert 1 candidate:
PKS 2004-447.  This Seyfert is consistent with the formal definition 
for NLS1s, although it does not have quite the same spectral features
as some typical members of this subclass.  Only ROSAT survey data is available at X-ray wavelengths, so it has not been possible to
compare this source with other NLS1s at these wavelengths. A full
comparison of this source with other members of the subclass will
improve our physical understanding of NLS1s. In addition,
using standard calculations, we estimate the central black hole to have
a mass of $\sim 5 \times 10^6 M_{\odot}$. This does not agree with
predictions in the literature, that radio-loud
AGN host very massive black holes.

\end{abstract}


\section{Introduction}
The radio-loud source PKS 2004-447, $z=0.24$, was identified as a candidate
Narrow-Line Seyfert 1 (NLS1) from low
resolution optical spectra of a subsample of quasars identified in the Parkes
Half-Jansky Flat-Spectrum Sample \citep{dri97}(hereafter PHFS). The
PHFS is designed to efficiently select radio-loud AGN
by looking for flat or inverted radio spectra over the range of 2.7
and 5.0 GHz. It contains 323 radio bright ($>0.5$Jy at 2.7 GHz),
generally compact radio sources. Higher resolution spectroscopy of
this object was obtained giving more accurate velocity widths of the
emission lines indicating this may be an NLS1 but more accurate
spectroscopy is still needed to be able to clearly separate the broad
and narrow components of H$\beta$. In any case this object has very
interesting implications to the study of AGN.

NLS1s are generally considered to be an extreme but common subclass of
AGN. They are defined by their optical emission line
properties, such that the H$\beta$ line is both strong (with a flux ratio
[\ion{O}{3}]/H$\beta<3$, similar to Seyfert 1 galaxies) and narrow (the 
H$\beta_{FWHM}<2000$km s$^{-1}$) \citep{ost85}. These properties
are correlated with strong \ion{Fe}{2} emission and a strong soft
X-ray excess, 
amongst other properties, for most objects in the subclass.
NLS1s do not appear to form a distinct class
but are instead connected to the ``standard'' broad-line Seyfert
population through a continuum of properties. \citet{bor92}
have shown that NLS1s cluster at one end of the region defined
by the first principal component described in their study of the
optical spectra of 87 low-redshift BQS quasars. The first
principal component is
thought to describe the major physical property in quasar
structure that is responsible for spectral differences between
AGN, independent of orientation. \citet{bor92}
suggest that the physical parameter driving this eigenvector is
$\dot{M} / \dot{M}_{Edd}$, where $\dot{M}$ is the rate of mass accretion onto the central massive object. NLS1s are thought to be accreting 
at a rate closer to the Eddington limit ($\dot{M}_{Edd}$) \citep{bor92,lao97}. 
NLS1s are generally radio-quiet objects, with only three
previously identified radio-loud objects, PKS 0558-504 \citep{rem86}, RGB
J0044+193 \citep{sie99} and J0134.2-4258 \citep{gru00}. 
\citet{bor92} find that
radio-loud QSOs and the NLS1s (with a strong, soft X-ray excess) lie at
opposite ends of the primary eigenvector.

Although the criteria for NLS1s are well defined, it is unclear whether these
phenomenological attributes reflect a single underlying physical mechanism.
Thus the discovery of a very radio-loud NLS1 may indicate that the
observational definition of NLS1s requires refinement. Alternatively, 
radio-loud NLS1s may 
provide a more stringent test of the models of NLS1s.
Three important consequences of the identification of a radio-loud NLS1
will be considered.  Firstly, are the radio-loud NLS1s the same
class of objects as most others in the class. Secondly, are radio-loud
NLS1s consistent with
any of the popular models for NLS1s and thirdly, do radio-loud quasars
require large mass black
holes. This object challenges that assertion as is further discussed in
section~4 

In section 2, we present the observational data on PKS 2004-477.
In section 3, this object is compared to the three other radio-loud NLS1s 
which have been identified. The central black hole mass is determined 
using standard techniques in section 4, and possible models for NLS1s 
are discussed in section
5.  Finally our conclusions are presented in section 6.

\section{Observational Data}

\subsection{Optical Spectrum}

PKS 2004-447 was first identified as a NLS1 candidate from a low
resolution spectrum obtained using the RGO/FORS spectrograph at the AAT in
1984 and published in \citet{dri97}.
A higher resolution confirmation spectrum
was taken using the double beam spectrograph (DBS) on the ANU 2.3m
telescope, 1st August 2000, and is shown in Fig~1. The conditions were
not photometric. The spectrum was reduced using standard
procedures in the IRAF. The spectrum has a resolution
of$\sim 2.2$\AA. 

\begin{figure}
\plotone{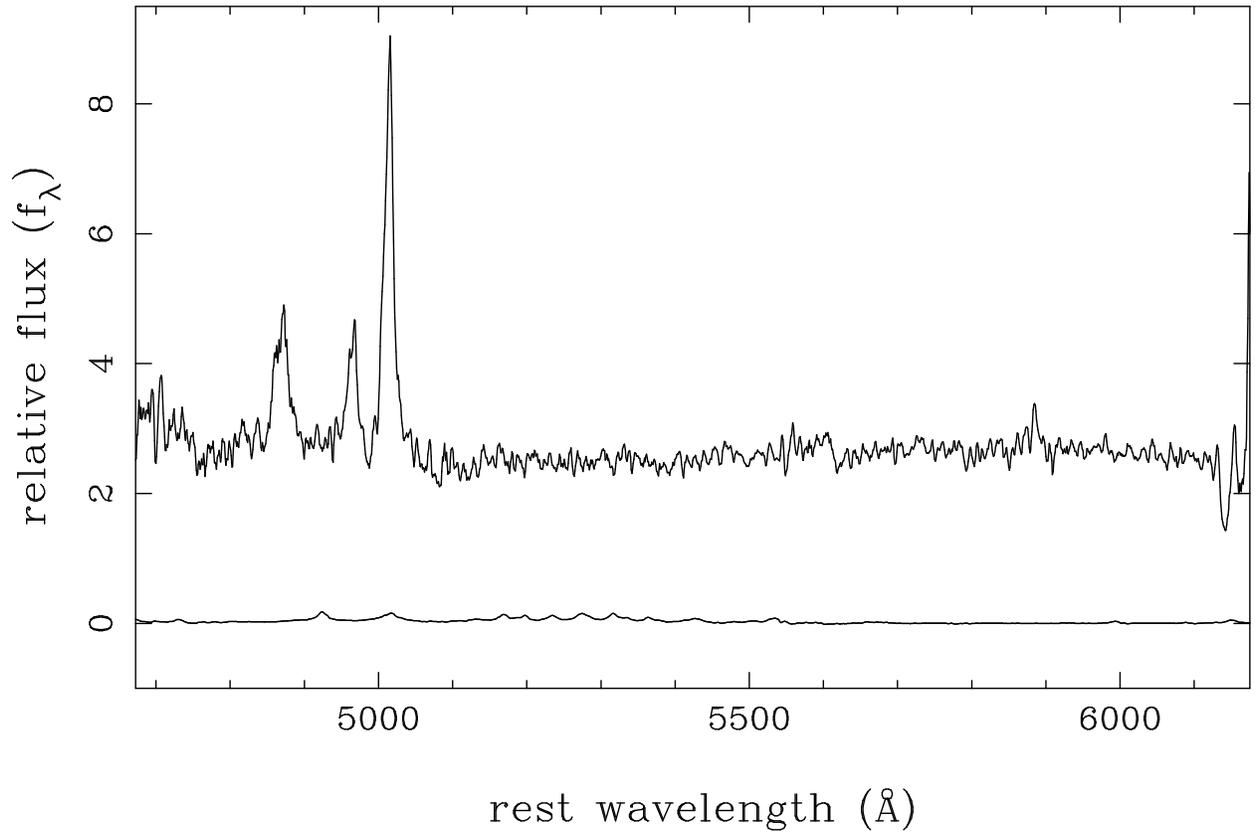}
\caption{Rest frame, medium resolution spectrum of PKS~2004-447 from the
DBS on the 2.3m telescope at the Siding Springs Observatory, showing the
H$\beta$ and [\ion{O}{3}] region (top) and the best-fit rescaled \ion{Fe}{2} spectrum shown
below it.}
\end{figure}

\begin{figure}
\plotone{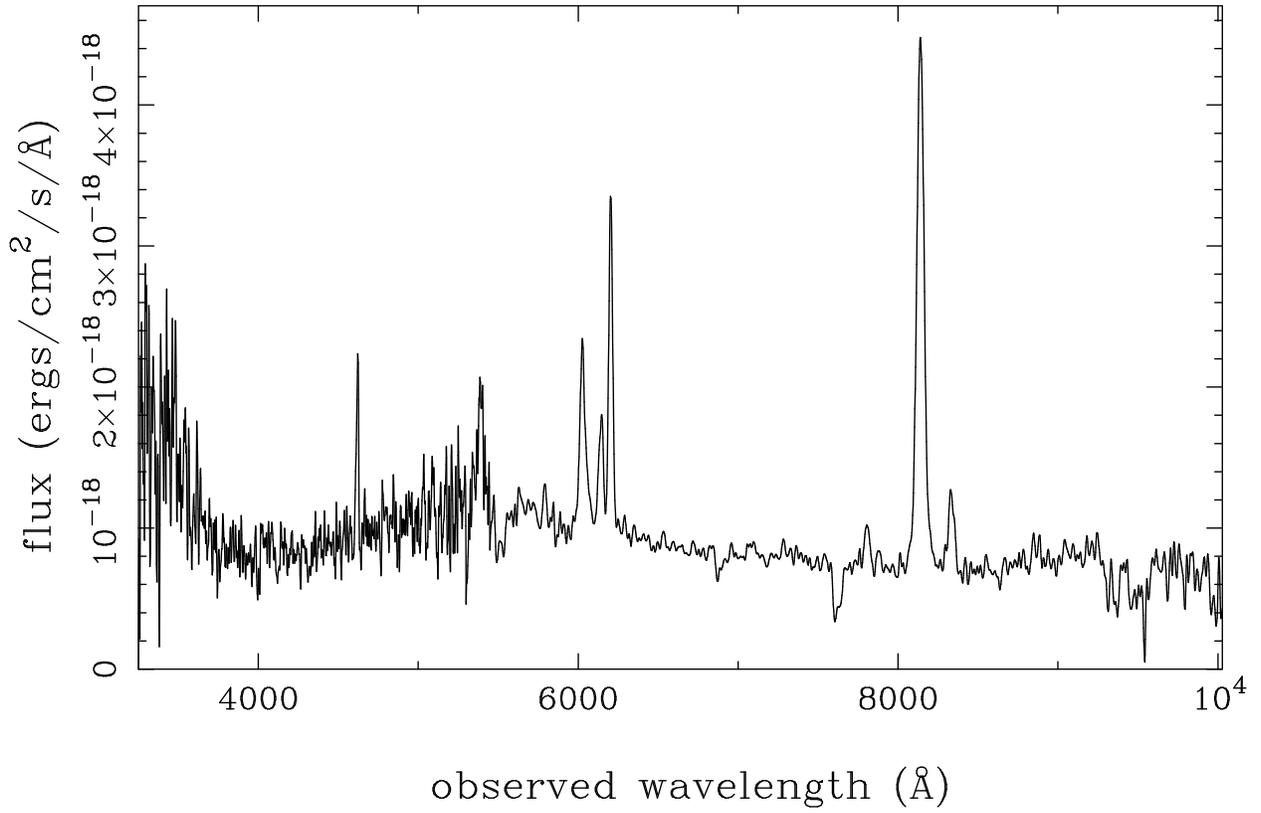}
\caption{Low resolution spectrum of PKS~2004-447 from RGO spectrograph
on the AAT.}
\end{figure}

Typically, in NLS1s, \ion{Fe}{2} emission contaminates the spectrum making
accurate measurements of H$\beta$ and [\ion{O}{3}] difficult. In order to
account for this and to get an estimate on the strength of the \ion{Fe}{2}
emission in this object we employed the \ion{Fe}{2} subtraction method
introduced by \citet{bor92} and now commonly used. We use a template
\ion{Fe}{2} spectrum taken from the prototype strong Fe emitter I Zw 1 and
scale it and shift it until the width and intensity match those seen
in our object. We did this over the region 5050-5500\AA~ using a
$\chi ^2$-minimisation to get the best value
for the scaling. We find that this spectral region contains very
little \ion{Fe}{2} emission with an
equivalent width EW$< 10$\AA~for the whole \ion{Fe}{2} complex in the region
5050-5450 \AA. Compared to values measured in \citet{bor92}, this
is extremely low (although we are fitting a slightly different
region of the \ion{Fe}{2} spectrum).  
We measured the line widths and fluxes using Lorentzian
fits to the emission lines and the results are summarised in Table 1. 
The width of $H\beta _{FWHM}=1447$km s$^{-1}$and the flux ratio
[\ion{O}{3}]/H$\beta= 1.6$,  
fit the criteria for
classification as a NLS1. The DBS spectrum does not extend to
wavelengths of the \ion{Fe}{2}($\lambda$4570) lines.  However the previous low
resolution spectrum indicates \ion{Fe}{2} emission in the region 4435-4700
\AA, blueward of the $H\beta$
line (Fig 2). The low resolution spectrum also shows strong H$\alpha$
and evidence for H$\gamma$ emission and it also indicates that the
stregth of the H$\beta$ emission, relative to [\ion{O}{3}] has varied
between the two epochs.

PKS 2004-447 is also optically variable. The magnitude of
this object measured from the COSMOS/UKST Southern Sky Catalogue is
$B_{J}=18.1$, while more recent photometry obtained
on the ANU 2.3m telescope gave $B=19.5$ \citep{fra00}, indicating a
drop in flux by a factor of $\approx 4$ over an interval of
several years. Simultaneous optical/IR photometry are shown in
Fig~3. The continuum is very red and there is no
evidence of a big blue bump.
The absolute magnitude is between -19.0 and -21.2, for $q_{0} =0.5$ and
$H_0 =100$, clearly placing this in the Seyfert luminosity class.

\begin{table}
\begin{center}
\caption{ Equivalent widths of the emission lines\tablenotemark{a}}
\begin{tabular}{|p{3.5cm}||p{2.5cm}|p{2.5cm}|p{2.5cm}|}
\hline & H$\beta$ (4861) & \ion{O}{3} (4959) & \ion{O}{3} (5007)\\
\hline
\hline FWHM(km s$^{-1}$) & 1447 & 951 & 754 \\
\hline flux relative to H$\beta$ & 1.0&0.61 &1.61 \\
\hline EW (\AA)& 31.6&19.6 &50.6 \\
\hline Offset (km s$^{-1}$)& $+28$&$-78$&0\\
\tableline
\end{tabular}
\tablenotetext{a}{Measured from the spectrum in Fig~1; details are described in the text.}
\end{center}
\end{table}

\subsection{Radio Emission}
Under the selection criteria, the PHFS objects are detected
above 0.5 Jy at 2.7GHz and with a spectral slope $\alpha < 0.5$
(F$_{\nu}\propto \nu^{-\alpha}$) taken from
non-simultaneous observations at 2.7GHz and 5.0 GHz. Using this method PKS
2004-447 had a 2.7 GHz flux of 0.81 Jy and a spectral index $\alpha_{r} =
0.36$. Subsequent simultaneous observations of this source were taken
using the ATCA (23rd November 1995) and the radio flux had
varied since the Parkes observations,
demonstrating long term variability at these
frequencies as well. The radio spectrum is shown in Fig~3. A powerlaw
fit to the simultaneous data gives a spectral index $\alpha_{r} = 0.67$;
thus PKS 2004-447 is actually a steep spectrum source. Calculating
$R$, where $R$ is the ratio of radio to optical flux ($f_{4.85GHz}/f_{B}$), \citep{kel89}, gives
values in the range $1710 < R < 6320$ depending on  value used for the
optical magnitude. Obviously, this source is very radio-loud.

\begin{figure}
\plotone{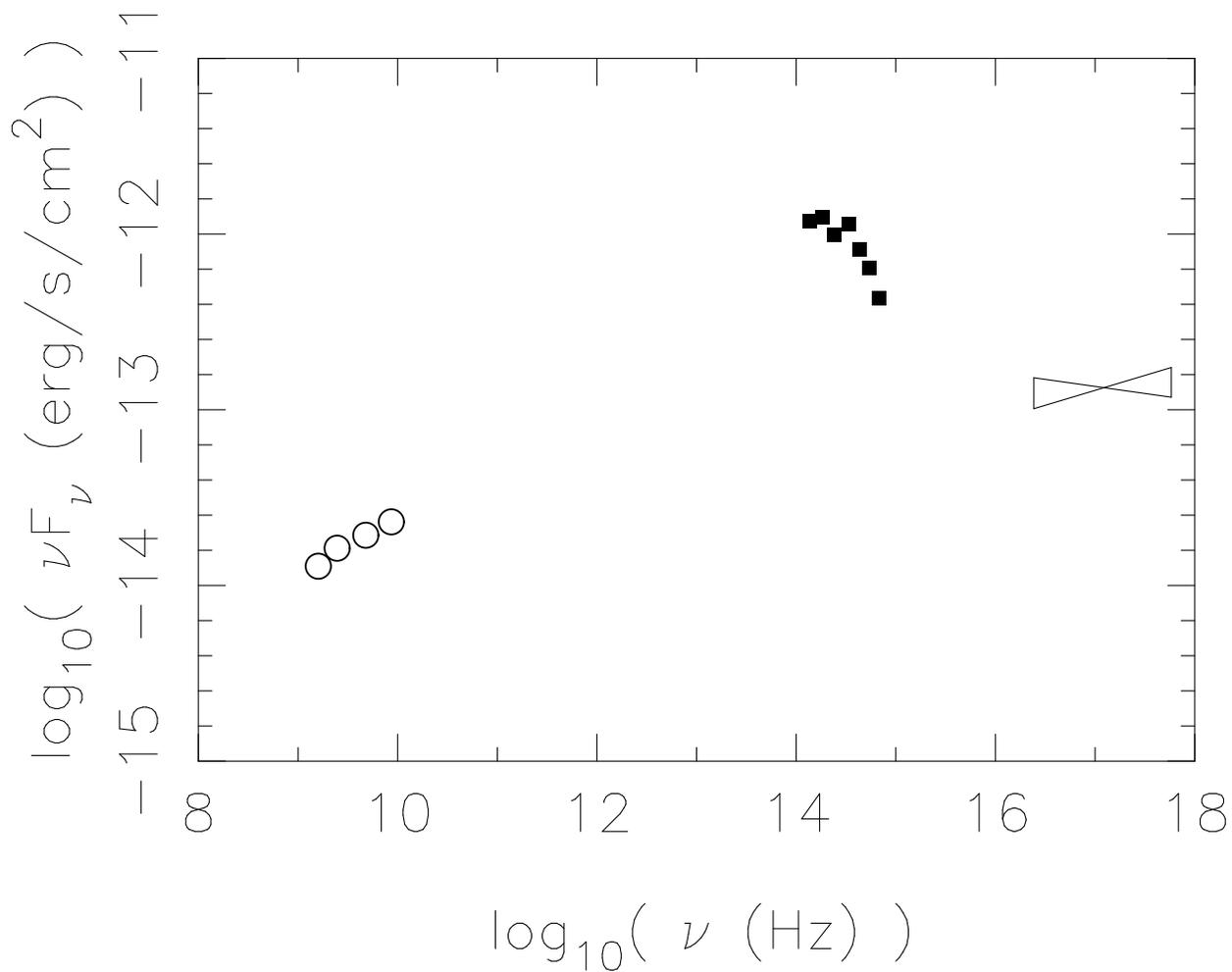}
\caption{Spectral energy distribution of PKS 2004-447. Radio data from
contemporaneous observations at the ATCA (circles, unpublished data). Optical data from
quasi-simultaneous observation at Siding Spring \citep[squares]{fra00}. X-ray data from ROSAT
All Sky Survey \citep[line]{sie98}.}
\end{figure}

The radio image is
unresolved in the ATCA observations.
The visibility of PKS 2004-477 was measured using the Parkes-Tidbinbilla Interferometer on a 275 km baseline at 2.3 GHz \citep{dun93}. The average
of two measurements gave a visibility of 0.680 corresponding to an
angular size of $\sim 0.036 \arcsec$ or $\sim 85 pc$. Since this source size
is derived from a 2-element interferometer, it is measured along one baseline.

There are two previous studies of the radio properties of NLS1s. The
most recent and well defined is that of \citet{mor00}, who obtained high
resolution observations of 24 NLS1s at 20cm and 3.6cm using the VLA.
All but one of the sources were detected, with 20cm fluxes up to $10^{25} W\,
Hz^{-1}$, which is higher, on average, than classical Seyferts.
Moran found most NLS1s unresolved at $\sim 1 \arcsec$, with 
generally quite steep ($\alpha \approx 1.1-1.2$) radio spectra, though the
spectral index of  PKS 2004-477 is consistent with the flattest
values. In other words these could be considered as low luminosity
compact steep spectrum (CSS) sources.
There is also a few cases of variability at 20cm, again,
as seen in PKS 2004-477. Thus PKS 2004-477 is compact, steep-spectrum 
and variable in the radio, which are similar properties to other NLS1s
\citet{mor00}.  This is further discussed in section 5.

\subsection{X-ray Emission}
NLS1s have unusual X-ray properties. They generally display a
large soft X-ray excess, a steep hard X-ray spectrum and large
amplitude X-ray variability over short and long timescales.
\citet{sie98} investigated the X-ray properties of the PHFS using the
ROSAT All Sky Survey. PKS 2004-447 was detected with a total
flux in the 0.1-2.4 keV range of $0.427 \pm 0.218 \times 10^{-12}$ erg
cm$^{-2}$ s$^{-1}$ corresponding to $L_X = 2.9 \times 10^{43} $erg
s$^{-1}$. We find an optical--to--X-ray slope of
$\alpha_{ox}=-1.2$. This is consistent with values found by
\citet{xu99} for a ROSAT selected sample of NLS1s.

\section{Previous Detections of Radio-Loud NLS1s}

Most NLS1s are radio-quiet and only three radio-loud NLS1s are 
known \citep{sie99}. 
PKS 0558-504 was optically identified on the basis of X-ray positions
from the High Energy Astronomical Observatory \citep{rem86}. 
It was noted as not only having very narrow hydrogen emission lines,
but also strong \ion{Fe}{2} emission.  It has a radio flux of 113 mJy at 4.85
GHz, an optical 
magnitude $m_B = 14.97$ and a redshift $z=0.137$. This gives  $R \approx 27$
\citep{sie99}. This source exhibits strong X-ray variability
on medium (months) and short (days, hours) time scales \citep{gli00}
and an X-ray flare indicating relativistic beaming \citep{rem91}.

\citet{sie99} identified RGB J0044+193 as a radio-loud NLS1.
RGB J0044+193 has only a moderate radio flux, but is
calculated to have a
radio-to-optical flux ratio $R \approx 31$. 
It was identified as a radio-loud X-ray
source with a redshift of $0.181$ in a cross-correlation of 
the ROSAT All-Sky Survey and the
Green Bank 5GHz radio survey \citep{lau98}. It has the bluest optical
spectrum of
any NLS1 observed to date. \citet{sie99} measure an optical continuum
slope of $\alpha = 1.3$ redward of 5000\AA\  and a slope of $-3.1$ blueward 
of 5000\AA\, where $f_{\nu} \propto \nu^{-\alpha}$. 
This source was detected in the 87GB
survey with $f_{4.85GHz}=24$ mJy. A high resolution follow-up using
the VLA at 4.85 GHz gave a flux of only 7 mJy \citep{lau97},
indicating that this 
source is extended or variable. The latter is supported by the fact
that the radio source is unresolved on the VLA map and it is not
detected at 1.4 GHz in the NRAO/VLA Sky Survey, which is sensitive down
to 2.5 mJy. This suggests that the classification of the source as radio-loud
is uncertain. In all other respects, RGB J0044+193 is indistinguishable from
radio-quiet NLS1s.

RX J0134.2-4258 was discovered by \citet{gru00} to have the steepest
soft X-ray spectrum observed during the ROSAT All-Sky Survey. It has an
optical magnitude of $m_V =16.2$ and $z=0.237$.  It was detected in the
PMN survey \citep{wri94} with a flux of 0.055 Jy at 4.85 GHz, 
and was subsequently re-observed at the VLA to have
a flux of 0.025 Jy at 8.4 GHz. This gives a ratio of $R=71$, and a
radio spectral index of $\alpha = 1.4$ for non-simultaneous observations.
The source is highly variable in the X-ray, and there is evidence for
variability at other wavelengths.  The \ion{Fe}{2} emission is strong, unlike
that of PKS 2004-477.

\section{Black Hole Mass}

There seems to be a growing consensus that a more
massive black hole is needed to produce a radio-loud quasar
\citep{lao00,kas00,mcl00,pet00}. PKS~2004-447 appears to 
directly conflict with this
statement. Calculations of the black hole mass rely on the assumption
that the dynamics of the BLR gas are dominated by the central black
hole. To calculate the black hole mass for this object we use the
results of \citet{kas00}, which are derived from a
reverberation study of 17 quasars. 
\citet{kas00} empirically determine linear relationship between
$R_{BLR}$ and the luminosity ($\lambda L_{\lambda}$) which we use to estimate
the radius of the BLR-emitting gas. From equation~5 in \citet{kas00}.
\begin{equation}
R_{BLR} = (32.9)\left[\frac{\lambda L_{\lambda}(5100
{\text \AA})}{10^{44} {\text erg~s}^{-1}}\right]^{0.700} \text{lt-days}
\end{equation}
We use the luminosity taken from a linear interpolation between
photometric data points \citep{fra00} at the rest wavelength of 5100\AA.
Following \citet{kas00}, the mass of the black hole is given by
$M_{BH}=rv^{2}G^{-1}$. To determine $v$, the velocity, we correct
$v_{FWHM}$ of the H$\beta$ emission line by a factor of $\sqrt{3}/2$ to account for velocities in
three dimensions. The mass is then
\begin{equation}
M = 1.464 \times 10^{5} \left(\frac{R_{BLR}}{\text{
lt-days}}\right)\left(\frac{v_{FWHM}}{10^{3}{\rm km~s}^{-1}}\right)^{2}
M_{\odot}
\end{equation} 
Using the cosmology $q_{0} =0.5$ and $H_{0}=100$km s$^{-1}$Mpc$^{-1}$,
this gives us a value for the central black
hole mass of $5.4 \times 10^6 M_{\odot}$. This
mass is two orders of magnitude lower than those obtained by other
authors for radio-loud quasars. In
a study of the black hole masses of quasars from the \citet{bor92}
sample, it was found that all quasars with $M_{BH}<3 \times 10^8
M_{\odot}$ are radio-quiet \citep{lao00}. The data from this sample is shown
in Fig~4, with the mass of PKS~2004-447 added. It can be seen that
PKS~2004-447 is
quite different than the rest of the sample. We note that
\citet{lao00} assumes a flatter relationship than \citet{kas00}, but
estimates that this difference results in a discrepancy of $<50\%$ in
the mass estimates. This corresponds to a shift of $<0.3$ in
log$M_{BH}$, which does not qualitatively affect the result in
figure~4. Similarly 
\citet{mcl00}, in their study of radio-loud and radio-quiet objects,
came to the conclusion that a radio-loud quasar requires a mass
$M_{BH}>6 \times 10^8 M_{\odot}$. This again,is more than two orders
of magnitude
greater than the mass we derive for PKS~2004-447. 

We speculate that
the reason our object has a black hole mass so much smaller than
those seen in previous studies is not entirely due to it being an
unusual object, but more likely due to the selection criteria in the
previous samples. All previous studies have used optical
selection to compile their sample. This selection technique
preferentially selects radio-loud
quasars which are blue with very broad emission lines. However samples of
radio-selected quasars such as the PHFS show that the general radio-loud
population can be quite different \citep{web95,fra00}. In these
samples it may be
quite common to have low mass black holes producing large radio-jets
and this issue will be addressed in a future paper.

\begin{figure}
\plotone{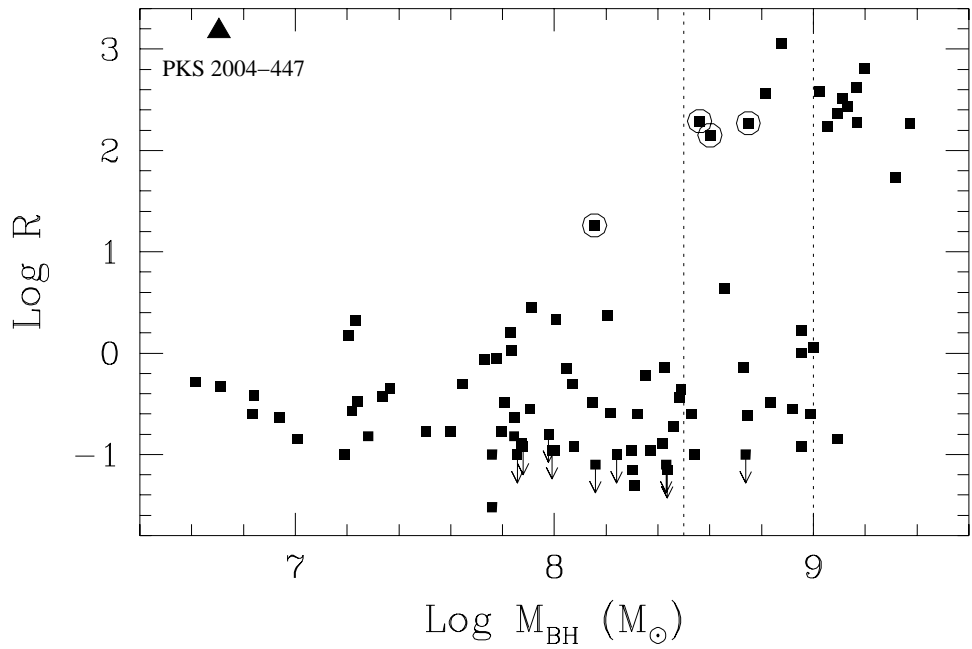}
\caption{Radio loudness vs. black hole mass, showing PKS 2004-447 in
relation to data from figure~2 of \citep{lao00}. The \citet{lao00}
data points are from the \citet{bor92} sample of 87 low redshift
quasars. The circled squares were proposed by \citet{fal96a,fal96b} to
be beamed intrinsically radio-quiet quasars. PKS~2004-447 has been
placed on the diagram at its lowest value for $R$ quoted in text.}
\end{figure}

\section{Discussion of Physical models}

\subsection{Classification}

Since the subclass of NLS1s are defined by phenomenological
measurements, rather than a specific physical model, it
is possible that more than one physical mechanism will produce
the defining parameters. First we should consider whether
PKS~2004-447 should be defined as an NLS1.  Certainly it meets
the formal definition of \citet{ost85}.  In the X-ray, the
flux ratios are consistent with other NLS1s, but pointed observations
are required to obtain spectral information.  In the optical,
the \ion{Fe}{2} emission is much weaker than other NLS1s. Apart from the
obvious strength of the radio
flux, the size of the radio source, and its spectral index are
consistent with other radio measurements \citep{mor00}. 

The object's radio properties fit with the definition of a Compact
Steep Spectrum radio source (CSS) according to the definition from \citet{fan95}
of: $P_{tot} > 10^{26}$W Hz$^{-1}$; a size
smaller or comparable to the optical galaxy scale; and the spectral
index $\alpha_{r} >0.5$. The CSS sources tend to have spectra with
narrow emission lines typical of radio-galaxies.
The broad component
of H$\beta$ seems to be extremely weak or absent
\citep{mor97}. \citet{gel94} find an average value [\ion{O}{3}]/H$\beta \sim
5$ indicating that the H$\beta$ flux is too weak to be an NLS1. In the study
of \citet{gel94} all CSS sources with measurable H$\beta$ are of
quasar luminosity. So PKS 2004-447 is unusual in both categories: firstly it has
an optical spectrum which is unusual for CSS source; and secondly has
radio power unusual for NLS1s.

The source is clearly in the Seyfert luminosity range, as are
a significant fraction of the sources in PHFS.  
The differences in optical spectral properties, (such as the weakness of
the \ion{Fe}{2} emission) between
PKS 2004-477 and NLS1s in general raise the possibility
that the NLS1-defining characteristics of this source are
due to a different physical mechanism
from other objects in the subclass.  This possibility may be
resolved when an X-ray spectrum is obtained.

\subsection{Physical Models}

Finding unusual objects in a particular class is a useful
method for distinguishing between different possible physical models. 
Although there have been a variety of
physical models suggested for NLS1s \citep[see for a summary]{tan99}
it seems that the  model with the strongest observational support
is the high accretion rate, low black hole mass model.
This model suggests that NLS1s are objects where the black
hole is accreting at a rate closer to the Eddington limit compared to
broad-line Seyferts. Since there is a higher rate of accretion,
a given luminosity corresponds to a black hole with a smaller mass. 
If the size of the BLR is determined by the bolometric luminosity, then,
at a given radius, the broad line clouds will have slower velocities 
and the emission lines will be narrower. 
This model was developed to explain the soft X-ray excess and the
steep X-ray powerlaw, by analogy with the X-ray spectra of galactic
black hole candidates accreting in their ``high''state \citep{pou95}. 
For high accretion rates, the soft thermal emission from the
disk becomes energetically dominant, producing the soft X-ray excess,
and the X-ray powerlaw spectrum can be steepened as a result of Compton
cooling of the electrons by the soft X-ray photons.

An alternative model for the narrow Hydrogen lines suggests we are
viewing the source near the axis of the accretion disk.  If the
velocity width of the Hydrogen emission lines is dominated by the
rotation of the accretion disk, then they will appear narrow.
The radio emission in PKS is moderately steep spectrum, which
weakens the argument that we observe this source near the axis.
Indeed it also weakens arguments which suggest that the radio
emission is strongly boosted by beaming.  However the radio
source is very compact, with no evidence at any frequency for
double-lobed emission. 

\section{Conclusion}

PKS 2004-447, detected in the PHFS, is an unusual source. To summarize the major points regarding PKS~2004-447:
\begin{enumerate}
\item It has H$\beta$ width and strength consistent with being
classified as an NLS1 as defined by \citet{ost85}.
\item Its radio properties are very unusual for a Seyfert Galaxy
having the following characteristics:
very strong radio flux, 0.81 Jy at 2.7 GHz;
very radio-loud, $R>1700$;
steep radio spectral index $\alpha_{r}=0.67$;
compact radio source;
some evidence of long term variability.
These properties are consistent with it being classified as a
Compact Steep Spectrum source.
\item It has been detected in the RASS with a flux $0.427 \times
10^{-12}$ erg~cm$^{-2}$~s$^{-1}$ but no spectral information is available.
\item Following the procedure set out by \citet{kas00}, we calculate
the black hole mass to be $5.4 \times 10^{6} M_{\odot}$. This mass is
more than 2 orders of magnitude lower than those seen previously for
radio-loud AGN and challenges previous results that a large black hole
mass is needed to produce radio-loud AGN.
\end{enumerate}
Additional observations of the X-ray
spectrum will further constrain models and may provide evidence that
the black hole is accreting at a rate closer to the Eddington
limit. This would be an indication that the underlying physical mechanism
in this object is similar to those in other NLS1s, and that
the radio power of the object is actually connected to a different
parameter.

We wish to thank Dirk Grupe for providing us with the
\ion{Fe}{2} template used in the analysis.


\begin{thebibliography}{}
%
\bibitem[Boroson and Green (1992)]{bor92} Boroson, T.\ A. and Green,
R.\ F. 1992 \apjs, 80, 109
%
\bibitem[Drinkwater et al.(1997)]{dri97} Drinkwater, M.\ J.,
    Webster, R.\ L., Francis, P.\ J., Condon, J.\ J., Ellison, S.\ L.,
    Jauncey, D.\ L., Lovell, J., Peterson, B.\ A., \&
    Savage, A. 1997, \mnras, 284, 85
%
\bibitem[Duncan et al. (1993)]{dun93} Duncan, R. A., White, G. L., Wark, R., Reynolds, J. E., Jauncey, D. L., Norris, R. P., Taaffe, L. and Savage, A.,
1993, PASA, 10, 310
%
\bibitem[Falcke, Patnaik, \& Sherwood(1996)]{fal96a} Falcke, 
H., Patnaik, A.\ R., \& Sherwood, W.\ 1996a, \apjl, 473, L13 
%
\bibitem[Falcke, Sherwood, \& Patnaik(1996)]{fal96b} Falcke, 
H., Sherwood, W., \& Patnaik, A.\ R.\ 1996b, \apj, 471, 106 
%
\bibitem[Fanti et al.(1995)]{fan95} Fanti, C., Fanti, R., 
Dallacasa, D., Schilizzi, R.\ T., Spencer, R.\ E., \& Stanghellini, C.\ 
1995, \aap, 302, 317 
%
\bibitem[Francis, Whiting, \& Webster(2000)]{fra00} Francis, 
P.\ J., Whiting, M.\ T., \& Webster, R.\ L.\ 2000, Publications of the 
Astronomical Society of Australia, 17, 56 
%
\bibitem[Gelderman and Whittle (1994)]{gel94} Gelderman, R., \&
Whittle, M., 1994, \apjs, 91, 491
%
\bibitem[Gliozzi et al.(2000)]{gli00} Gliozzi, M., Boller, Th., Brinkmann, W.
and Brandt, W. N., 2000, \aap, 356, 17
%
\bibitem[Grupe, Leighly, Thomas, \& 
Laurent-Muehleisen(2000)]{gru00} Grupe, D., Leighly, K.\ M., 
Thomas, H.\ -., \& Laurent-Muehleisen, S.\ A.\ 2000, \aap, 356, 11 
%
\bibitem[Kaspi et al.(2000)]{kas00} Kaspi, S., Smith, P.\ S., 
Netzer, H., Maoz, D., Jannuzi, B.\ T., \& Giveon, U.\ 2000, \apj, 533, 631 
%
\bibitem[Kellerman et al.(1989)]{kel89} Kellerman, K. I., Sramek, R.,
Schmidt, M., Shaffer, D. B., Green, R., 1989, \aj, 98, 1195
%
\bibitem[Laor et al.(1997)]{lao97} Laor, A., Fiore, F., 
Elvis, M., Wilkes, B.\ J., \& McDowell, J.\ C.\ 1997, \apj, 477, 93 
%
\bibitem[Laor(2000)]{lao00} Laor, A.\ 2000, \apjl, 543, L111
%
\bibitem[Laurent-Muehleisen et al.(1997)]{lau97} 
Laurent-Muehleisen, S.\ A., Kollgaard, R.\ I., Ryan, P.\ J., Feigelson, E.\ 
D., Brinkmann, W., \& Siebert, J.\ 1997, \aaps, 122, 235 
%
\bibitem[Laurent-Muehleisen et al.(1998)]{lau98} 
Laurent-Muehleisen, S.\ A., Kollgaard, R.\ I., Ciardullo, R., Feigelson, 
E.\ D., Brinkmann, W., \& Siebert, J.\ 1998, \apjs, 118, 127 
%
\bibitem[Moran (2000)]{mor00} Moran, E, C., 2000, NewAR, 44, 527
%
\bibitem[Morganti (1997)]{mor97} Morganti, R., Tadhunter, C.\ N.,
Dickson, R., Shaw, M., \aap, 326, 130
%
\bibitem[McLure and Dunlop (2000)]{mcl00} McLure, R.\ J. \& Dunlop, J.\
S., astro-ph/0009406.
%
\bibitem[Osterbrock and Pogge (1985)]{ost85} Osterbrock, D. E. and
Pogge, R. W., 1985, \apj, 297, 166
%
\bibitem[Peterson et al.(2000)]{pet00} Peterson, B.\ M.\ et 
al.\ 2000, \apj, 542, 161 
%
\bibitem[Pounds et al.(1995)]{pou95} Pounds, K. A., Done, C. and Osborne, J. P.,
1995 \mnras, 277, L5
%
\bibitem[Remillard et al.(1986)]{rem86} Remillard, R. A., Bradt, H. V., Buckley, D. A. H., Roberts, W., Schwartz., D. A., Tuohy, I. R \& Wood, K. 1986, \apj,
301, 742
%
\bibitem[Remillard et al.(1991)]{rem91} Remillard, R. A., Grossan, B., 
Bradt, H. V., Ohashi, T. \& Hayashida, K., 1991 \nat, 350, 589
%
\bibitem[Siebert et al.(1998)]{sie98} Siebert, J., Brinkman, W., Drinkwater, M. J.,
Yuan, W., Francis, P. J., Peterson, B. A. \& Webster, R. L. 1998, \mnras, 
301, 261
%
\bibitem[Siebert et al.(1999)]{sie99} Siebert, J., Leighly, K. M.,
Laurent-Muehleisen, S. A., Brinkman, W., Boller, \& Matsuoka, M.,  1999,
\aap, 348, 678
%
\bibitem[Taniguchi et al.(1999)]{tan99} Taniguchi, Y., Murayama, T., \&
Nagao, T., 1999, astro-ph/9910036
%
\bibitem[Webster et al.(1995)]{web95} Webster, R.\ L., 
Francis, P.\ J., Peterson, B.\ A., Drinkwater, M.\ J., \& Masci, F.\ J.\ 
1995, \nat, 375, 469 
%
\bibitem[Wright et al.(1994)]{wri94} Wright, A. E., Griffith, M. R.,
Burke, B. F. \& Ekers, R. D., 1994, \apjs, 91, 111
%
\bibitem[Xu, Wei \& Hu (1999)]{xu99} Xu, D.\ W., Wei, J.\ Y. \& Hu, J.\ Y.,
1999, \apj, 517, 622
%
\end{thebibliography}
\end{document}